\documentclass[useAMS,usenatbib,twcolumn,usegraphicx]{mn2e}

\newcommand{\beq}{\begin{equation}}
\newcommand{\eeq}{\end{equation}}
\newcommand{\beqar}{\begin{eqnarray}}
\newcommand{\eeqar}{\end{eqnarray}}

\title[Constraints on Pasta Structure of Neutron Stars]
{Constraints on Pasta Structure of Neutron Stars from Oscillations in Giant Flares}
\author[H. Sotani]
{Hajime Sotani \thanks{E-mail:hajime.sotani@nao.ac.jp}
\\
Division of Theoretical Astronomy, National Astronomical Observatory of Japan, 
2-21-1 Osawa, Mitaka, Tokyo 181-8588, Japan}
\begin{document}

\maketitle

\label{firstpage}

\begin{abstract}
  We show that the shear modes in the neutron star crust are quite sensitive to the existence of nonuniform nuclear structures, so-called ``pasta". Due to the existence of pasta phase, the frequencies of shear modes are reduced, where the dependence of fundamental frequency is different from that of overtones. Since the torsional shear frequencies depend strongly on the structure of pasta phase, through the observations of stellar oscillations, one can probe the pasta structure in the crust, although that is quite difficult via the other observations. Additionally, considering the effect of pasta phase, we show the possibility to explain the all observed frequencies in the SGR 1806-20 with using only crust torsional shear modes.
\end{abstract}


The soft gamma repeaters (SGRs) are considered as one of the most promising candidate of magnetars, which are neutron stars with strong magnetic fields \citep{DT1992}. The sporadic X- and gamma-ray bursts are radiated from SGRs, while SGRs rarely emit much stronger gamma-rays 
called ``giant flares". Up to now, at least three giant flares have been detected, which are the SGR 0526-66 in 1979, the SGR 1900+14 in 1998, and the SGR 1806-20 in 2004. For each giant flare phenomenon, one can observe a decaying softer part (tail) for hundreds of seconds after the initial short peak in the hard part of the spectrum. Through the timing analysis of these decaying tail, the quasi-periodic oscillations (QPOs) have discovered, which are in the range from tens Hz up to a few kHz \citep{Israel2005,WS2006}. Since the QPOs are believed as the outcomes of the neutron star oscillations, the observations of QPOs in SGRs could be first evidences to detect the neutron star oscillations directly.

The current understanding about the observed QPO frequencies in SGRs is as follows; a part of lower frequencies such as 18, 26, and 30 Hz in the SGR 1806-20 are associated with the magnetized fluid core, while the others are interpreted as torsional shear modes of the solid crust. However, it seems to be more complicated to understand the oscillations of magnetized neutron stars completely. The oscillations in the core become the Alfv\'{e}n continuum \citep{Levin2006,Levin2007,Sotani2008a,CBK2009,CSF2009}, and one might need to consider the coupling between oscillations in the fluid core and in solid crust if the magnetic field is stronger \citep{Levin2006,Levin2007}. Additionally, \cite{vHL2011} described how crustal modes may survive in the gapes left in the Alfv\'{e}n continuum. In the spite of these complexities, the crustal torsional modes can still emerge globally if the magnetic field is not so strong, e.g., $B<10^{14}$ G, and the frequencies of torsional shear modes are almost same values as those in the case without magnetic filed \citep{GCFMS2011,CK2011}. Thus, comparing the analysis of shear modes with the observed QPO frequencies, one can know the proper properties of the crust in neutron star. In fact, via the observed QPO frequencies, it could be possible to constrain on the stellar properties \citep{SA2007} and the distribution of stellar magnetic field \citep{Sotani2008b}.

It is considered that the neutron star crust exists from the bottom of the ocean of melted iron at a density $\sim10^6-10^8$ g/cm$^3$ inward to the boundary with the inner fluid core at a density of order the saturation density of nuclear matter $\rho_s\sim 3\times 10^{14}$ g/cm$^3$. Although nuclei in the crust form a bcc lattice due to Coulomb interactions, according to the recent studies, the nuclear structure in the bottom of crust could be nonuniform, i.e., with increasing the density, the shape of nuclear matter region is changing from sphere (bcc lattice) into cylinder, slab, cylindrical hole, and uniform matter (inner fluid core) \citep{LRP1993,O1993,SOT1995}. This variation of nuclear structure is known as the so-called ``pasta structure". The density that the cylinder structure appears, $\rho_p$, depends on the nuclear symmetry energy expressed with the density symmetry coefficient $L$ \citep{OI2007}, which is suggested to be order $\rho_p\sim10^{13}$ g/cm$^3$ via the calculations of the ground state of matter in the crust \citep{LRP1993,O1993,SOT1995}. However, it might be quite difficult to verify the existence of pasta structure by using the observation of neutron star properties such as mass and radius, because the width of pasta phase is around $10\%$ of the crust, which is less than a few hundred meters \citep{LRP1993}. In contrast, in this letter, we will calculate the torsional oscillations of neutron star with pasta phase and show that the frequencies of shear modes depend strongly on the presence of pasta phase and on $\rho_p$.

\begin{figure}
\begin{center}
\includegraphics[scale=0.45]{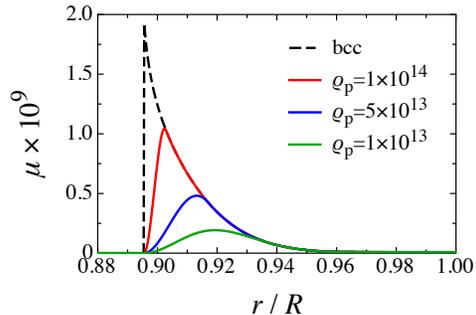} 
\end{center}
\caption{
Shear modulus $\mu$ as a function of relative radius $r/R$. Solid lines correspond to the shear modulus including pasta phase with different values of $\rho_p$, while broken line is corresponding to that shown by Eq. (\ref{eq:shear}), where the stellar mass is fixed that $M=1.4M_\odot$.
}
\label{fig:shear}
\end{figure}

Previously, there are many calculations about the shear modes (e.g., \cite{Lee2007,Sotani2007,SW2009}), where they assume that nuclei form bcc lattice in the crust. With this assumption, the shear modulus of the crust is suggested as
\begin{equation}
  \mu = 0.1194\times n_i(Ze)^2/a, \label{eq:shear}
\end{equation}
where $n_i$ is the ion number density, $a=(3/4\pi n_i)^{1/3}$ is the average ion spacing, and $+Ze$ is the ion charge \citep{SHOII1991}. Roughly speaking, this relation can be expressed as the power low with respect to the density \citep{Sotani2007}. On the other hand, it is pointed out that the elastic properties in the pasta phase could be liquid crystals rather than a crystalline solid \citep{PP1998}. That is, unlike the case of bcc lattice, the shear modulus should be decreasing in the pasta phase as increasing the density. This picture is thought to be natural, because the structure of nuclear matter changes gradually as mentioned the above and at last the shear modulus becomes zero in the fluid core. In order to realize such a relation about the shear modulus, we adopt Eq. (\ref{eq:shear}) in the crust except for the pasta phase, while in the pasta phase it is assumed that the shear modulus can be expressed as the cubic function with respect to the density, which satisfies that $\mu$ should connect to Eq. (\ref{eq:shear}) smoothly at $\rho=\rho_p$ and become zero smoothly at the boundary with the core, i.e., $\mu=c_1(\rho-\rho_c)^2(\rho-c_2)$, where $c_1$ and $c_2$ are some constants determined by the boundary conditions (see Fig. \ref{fig:shear}). Of course, this simple relation of $\mu$ in the pasta phase might be a kind of toy model, although expressed the rough behavior, because that should depend on the microscopic structure including the matter composition and/or the nuclear symmetry energy \citep{OI2007}. But still, this relation is thought to be enough to examine the dependence of shear oscillations on the existence of pasta phase as a first step. Anyway, as a result of the fall-off of shear modulus, one can expect that the shear velocity, $v_s=(\mu/\rho)^{1/2}$, will become smaller, and that the frequencies of shear modes will also decrease. In practice, although the understanding about the shear modulus of the liquid crystals is quite poor except for the suggestion that the shear modulus becomes small in the pasta phase \citep{PP1998}, it could be possible to obtain the information about the pasta phase and/or properties in liquid crystals via the observations of oscillations of neutron stars. Additionally, it should be noticed that the condition of $\mu=0$ at the crust-core interface corresponds to the limit of zero shear speed. Thus, the low-frequency modes might get trapped in the vicinity of crust-core interface and/or the large mode amplitude might produce at the interface, which may in turn imply a large boundary-layer damping.

%


To examine the shear modes, we prepare the static, spherically symmetric stellar models, which is the solution of the well-known Tolman-Oppenheimer-Volkoff (TOV) equations described by a metric of the form
\begin{equation}
  ds^2 = -e^{2\Phi}dt^2 + e^{2\Lambda}dr^2+r^2(d\theta^2 +\sin^2\theta d\phi^2),
\end{equation}
where $\Phi$ and $\Lambda$ are the function of the Schwarzshild radial coordinate $r$. Since the stiff equation of state (EOS) is favorable to the explanation of the observed QPO frequencies \citep{Sotani2007}, we adopt the EOS L \citep{EOSL} for the inner core. On the other hand, a modern EOS suggested in \citep{EOSDH} is adopted for the crust (see \cite{Sotani2007} for the stellar properties), where the density at the boundary with core is set to be $\rho_c=1.24\times 10^{14}$ g/cm$^3$ \citep{EOSDH}. Then, the unknown parameter to determine the stellar model is only the density that the pasta phase appears, $\rho_p$.

Since the pure torsional shear modes are incompressible and do not induce density variations in spherical stars, one can expect that no significant variation in the radiative part of the metric describing the gravitational field. Therefore, the frequencies of torsional shear modes are determined with satisfactory accuracy even when neglecting entirely the metric perturbations by setting $\delta g_{\mu\nu}=0$, i.e., adopting the relativistic Cowling approximation, and we adopt this approximation in this letter. The perturbation equation is derived from the linearized equation of motion, where one needs to provide the linearized shear stress tensor, $\delta T_{\mu\nu}^{(s)}$. In the same way as in \cite{ST1983}, in this letter, $\delta T_{\mu\nu}^{(s)}$ is assumed to be related to the linearized shear tensor, $\delta S_{\mu\nu}$, through $\delta T_{\mu\nu}^{(s)}=-2\mu \delta S_{\mu\nu}$. $\delta S_{\mu\nu}$ is determined from the relationship $\delta \sigma_{\mu\nu} = {\cal L}_{u}\delta S_{\mu\nu}$, where $\sigma_{\mu\nu}$ denotes the rate of shear tensor \citep{ST1983}.

The torsional shear modes can be described with using one perturbation variable, which is the angular displacement of the stellar matter, ${\cal Y}(t,r)$. The non-zero component of perturbed matter quantities is $\phi$-component of the perturbed 4-velocity of fluid, $\delta u^{\phi}$,  which is expressed with ${\cal Y}$ as
\begin{equation}
  \delta u^\phi = e^{-\Phi}\partial_t{\cal Y}(t,r)\frac{1}{\sin\theta}\partial_\theta P_{\ell}(\cos\theta),
\end{equation}
where $\partial_t$ and $\partial_\theta$ denote the partial derivative with respect to $t$ and $\theta$, respectively, while $P_\ell(\cos\theta)$ is the Legendre polynomical of order $\ell$. Assuming that the perturbed variable has a harmonic time dependence, such that ${\cal Y}(t,r)=e^{i\omega t}{\cal Y}(r)$, the perturbation equation reduces to
\begin{eqnarray}
  {\cal Y}'' &+& \left[\left(\frac{4}{r}+\Phi'-\Lambda'\right)+\frac{\mu'}{\mu}\right]{\cal Y}'  \nonumber \\
    &+& \left[\frac{\epsilon+p}{\mu}\omega^2e^{-2\Phi}-\frac{(\ell+2)(\ell-1)}{r^2}\right]e^{2\Lambda}{\cal Y} = 0,
\end{eqnarray}
where $\epsilon$ and $p$ correspond to the energy density and pressure, respectively, and the prime denotes the derivative with respect to $r$ \citep{ST1983}. With appropriate boundary conditions, the problem to solve becomes the eigenvalue problem. We impose a zero traction condition at the boundary between the inner core and the crust ($r=R_c$), and the zero-torque condition at the stellar surface $(r=R)$, which correspond to $\mu{\cal Y}'=0$ at $r=R_c$ and ${\cal Y}'=0$ at $r=R$ \citep{ST1983,Sotani2007}. So, if the shear modulus is exactly zero, the boundary condition at $r=R_c$ is automatically satisfied for arbitrary value of ${\cal Y}'$ and the number of boundary conditions is not enough to solve the problem. However, since the shear modulus in the realistic case could be not absolutely zero but quite small value with the some fluctuation, in this letter we adopt that ${\cal Y}'=0$ even at $r=R_c$.

\begin{figure}
\begin{center}
\includegraphics[scale=0.45]{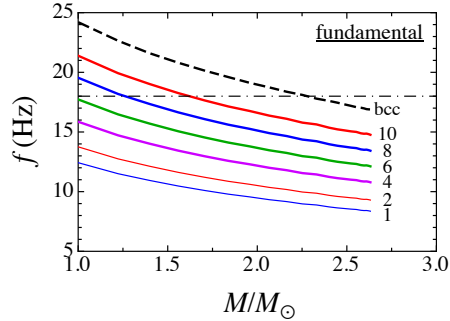} 
\end{center}
\caption{
Frequencies of fundamental torsional shear modes with $\ell=2$ as a function of neutron star mass. The broken line corresponds to the frequencies for the stellar model without pasta phase, while the solid lines correspond to those with pasta phase, where the labels denote the value of $\rho_p/(10^{13}$ g/cm$^3$). Additionally, the dot-dash-line denotes the lowest observed frequency in the SGR 1806-20, which is 18 Hz.
}
\label{fig:0f2}
\end{figure}

We examine the frequencies of torsional shear modes, as varying the value of $\rho_p$ in the range of $\rho_p=10^{13}-10^{14}$ g/cm$^3$, because that value is not certain but suggested to be order $10^{13}$ g/cm$^3$ as mentioned before \citep{LRP1993,O1993,SOT1995}. Fig. \ref{fig:0f2} shows the fundamental frequencies of torsional shear modes with $\ell=2$ as a function of the stellar mass, where the broken line corresponds to the frequencies for the stellar model without pasta phase and the solid lines are corresponding to those with pasta phase with different values of $\rho_p$. Obviously, one can observe that the frequencies of shear modes depend strongly on the existence of pasta phase. In fact, compared with the frequency for the stellar model without pasta phase, those with pasta phase are reduced to $12.0\%$, $34.9\%$, and $49.3\%$ for $\rho_p=1\times 10^{14}$, $4\times 10^{13}$, and $1\times 10^{13}$ g/cm$^3$, respectively. To compare with the observed frequencies, the lowest observed QPO frequency in the SGR 1806-20 (18Hz) is also plotted in this figure with the dot-dash line. From this figure, it is found that the smaller stellar mass is favored for smaller $\rho_p$, if the frequency of 18 Hz is explained as usual with the fundamental $\ell=2$ shear modes. Otherwise, with smaller $\rho_p$ than $6\times 10^{13}$ g/cm$^3$, one can explain the frequency of 18 Hz by using the fundamental shear mode with higher $\ell$.

\begin{figure}
\begin{center}
\includegraphics[scale=0.45]{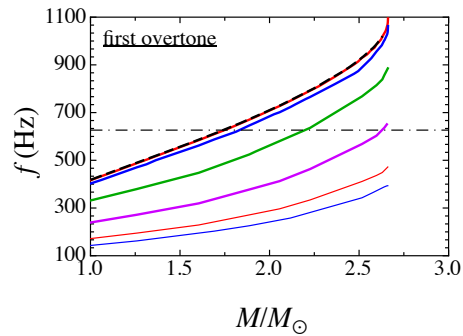} 
\end{center}
\caption{
Frequencies of first overtones of torsional shear modes with $\ell=2$ as a function of neutron star mass, where the meaning of lines are same as in Fig. \ref{fig:0f2}. The observed frequency in the SGR 1806-20, which is 626.5 Hz, is also shown with the dot-dash-line .
}
\label{fig:1f2}
\end{figure}

In Fig. \ref{fig:1f2}, the frequencies of 1st overtones of shear modes with $\ell=2$ are plotted as a function of the stellar mass. Unlike the fundamental modes, the frequencies with higher $\rho_p$ are almost same as that without pasta phase. This result might be corresponding to the tendency in the Newtonian limit, i.e., the frequency of fundamental modes could be roughly proportional to the shear speed, while those of overtone depend on the ratio between curst thickness and stellar radius, which are almost independent of the shear modulus \citep{HC1980}. Still, one can see the dependence of the frequency on the existence of pasta phase with lower $\rho_p$. In fact, the frequencies with pasta phase are different from that without pasta phase in $0.4\%$, $46.0\%$, and $67.0\%$ for $\rho_p=1\times 10^{14}$, $4\times 10^{13}$, and $1\times 10^{13}$ g/cm$^3$, respectively. This dependence might be difficult to explain in analogy with the Newtonian limit, but one could be possible to obtain the additional information of the crust property via the observation of frequencies of overtones. Similar to Fig. \ref{fig:0f2}, the observed frequency in the SGR 1806-20 (626.5Hz) is plotted in this figure with the dot-dash line, which is usually considered to be caused by the 1st overtone of shear modes \citep{Piro2005}. But, with smaller $\rho_p$, the frequency of 626.5 Hz might be corresponding to the 2nd or 3rd overtones.

%

Since the both frequencies of fundamental and overtone shear modes depend strongly on the presence of pasta phase, whose effect has been neglected so far, one needs to consider this effect on the shear oscillations. Additionally, owing to this strong dependence, it could be possible to probe the properties of pasta phase via the observations of stellar oscillations and stellar mass, although the constraint on the pasta phase is quite difficult via the other observations of neutron stars. In contrast to simple relation of $\mu$ adopted in this letter, we will make an examination with more realistic shear model in the pasta phase in the future.

\begin{figure}
\begin{center}
\includegraphics[scale=0.45]{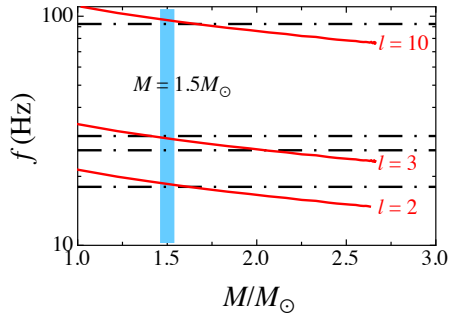} 
\end{center}
\caption{
Comparison of the frequencies of torsional shear modes (solid lines) with the observed QPO frequencies in SGR 1806-20 (dot-dash-lines), where the adopted stellar model is with $\rho_p=1\times 10^{14}$ g/cm$^3$.
}
\label{fig:fit114}
\end{figure}

At last, we will compare the calculated frequencies of shear modes with the observed QPO frequencies in giant flare. Especially, in this letter we focus on the QPO frequencies in the SGR 1806-20, i.e., 18, 26, 30, and 92.5 Hz in less than 100 Hz, because this phenomenon has the most observed frequencies among the detected giant flares in the past \citep{WS2006}. As mentioned in the introduction, some of the observed QPO frequencies are considered as a result of the crust shear modes, while the others are associated with the magnetized fluid core. The reason is because of the difficulty to explain the all observed frequencies with using only crust shear modes \citep{Sotani2007}. That is, the  fundamental $\ell=2$ mode, which is the possible lowest frequency, is considered to correspond to the observed frequency of 18 Hz and the fundamental $\ell=3$ mode is corresponding to 26 or 30 Hz. But, the spacing of shear frequencies with different $\ell$ is larger than the spacing between the observed frequencies of 26 and 30 Hz. In practice, according to such a traditional identification, we can explain the observed frequencies with the crust shear modes with pasta phase as shown in Fig. \ref{fig:fit114}, i.e., 18, 30, 92.5 Hz can be identified as $\ell=2$, $3$, and 10 fundamental modes within a few percent accuracy, where the expected stellar mass is $M=1.5M_\odot$.

\begin{figure}
\begin{center}
\includegraphics[scale=0.45]{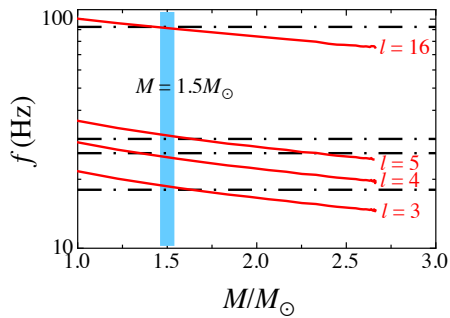} 
\end{center}
\caption{
Similar to Fig. \ref{fig:fit114}, but with $\rho_p=2\times 10^{13}$ g/cm$^3$.
}
\label{fig:fit213}
\end{figure}

However, we find the possibility to explain the all observed frequencies with using only crust shear modes, if $\rho_p$ would be small. As shown in Fig. \ref{fig:fit213}, the observed frequencies of 18, 26, 30, and 92.5 Hz can be identified as $\ell=3$, 4, 5, and 16 fundamental shear modes within a few percent accuracy again, where the expected stellar mass is $M=1.5M_\odot$. This is important suggestion to explain the observed QPO frequencies in giant flares, which could become a directing post in the asteroseismology with neutron stars.


In this letter, we consider the effect of nonuniform nuclear structure, so-called ``pasta", in the neutron star crust on the torsional shear modes. Based on the suggestion that the elastic properties in the pasta phase could be liquid crystals, the frequencies of shear modes are calculated with simple relation of shear modulus. As a result, due to the existence of pasta phase, one can observe the smaller frequencies than those expected without pasta phase. This result indicates not only the importance to take into account the pasta phase, but also the possibility to probe the pasta structure via the observations of the stellar oscillations, such as the QPO frequencies in giant flares. Furthermore, we show the possibility to explain the observed QPO frequencies with using only crust shear modes with pasta phase, which is quite difficult with the traditional identification without pasta phase. In order to find the most suitable stellar model with the observations, we need to adopt more realistic pasta structure and to examine systematically with different EOSs, where the additional effects, such as superfluidity inside the star \citep{CC2006,PA2011} as well as the stellar magnetic field, should be also taken into account. Still, we believe that the pasta structure in the neutron star curst could play an important role in the stellar oscillations.

At the end, we are grateful to T. Tatsumi, K. Iida, K.I. Nakazato, N. Yasutake for their fruitful discussions, and also to reviewer for careful reading and giving the valuable comments. This work was supported by Grant-in-Aid for Scientific Research on Innovative Areas (23105711).


\end{document}